\documentclass[twocolumn,showpacs,preprintnumbers,amsmath,amssymb]{revtex4}

\usepackage{graphicx}
\usepackage{dcolumn}
\usepackage{bm}

\begin{document}

\title{Enhance synchronizability by structural perturbations}
\author{Ming Zhao$^*$}
\author{Tao Zhou\footnote{They contributed equally to this work.}}
\author{Bing-Hong Wang}
\email{bhwang@ustc.edu.cn}
\author{Wen-Xu Wang}
\affiliation{%
Department of Modern Physics and Nonlinear Science Center,\\
University of Science and Technology of China, \\
Hefei Anhui, 230026, PR China
}%

\date{\today}

\begin{abstract}
In this paper, we investigate the collective synchronization of
system of coupled oscillators on Barab\'{a}si-Albert scale-free
network. We propose an approach of structural perturbations aiming
at those nodes with maximal betweenness. This method can markedly
enhance the network synchronizability, and is easy to be realized.
The simulation results show that the eigenratio will sharply
decrease to its half when only 0.6\% of those hub nodes are under
3-division processes when network size $N=2000$. In addition, the
present study also provides a theoretical evidence that the
maximal betweenness plays a main role in network synchronization.
\end{abstract}


\maketitle

\section{Introduction}
Many social, biological and communication systems can be properly
described as complex networks with nodes representing individuals
or organizations and edges mimicking the interactions among
them\cite{Reviews1,Reviews2,Reviews3}. One of the ultimate goals
of the current studies on topological structure of networks is to
understand and explain the workings of systems built upon those
networks: for instance, to understand how the topology of Internet
affects the spread of computer
virus\cite{Epidemic1,Epidemic2,Epidemic3}, how the structure of
power grids affects the cascading
behavior\cite{Cascade1,Cascade2,Cascade3}, how the connecting
pattern of intercommunication networks affects traffic
dynamics\cite{Traffic1,Traffic2,Traffic3}, and so on.

In the past few years, with the computerization of data
acquisition process and the availability of high computing powers,
scientists have found that many real-life networks share some
common statistic characteristics, the most important of which are
called small-world effect\cite{SWN1,SWN3} and scale-free
property\cite{SFN2}. The recognition of small-world effect
involves two facts, small average distance\cite{ex1}(as $L \sim
\texttt{ln}N$, where $N$ is the number of nodes in the network)
and large clustering coefficient\cite{ex2}. The number of edges
incident from a node $x$ is called the degree of $x$. The
scale-free property means the degree distribution of the network
obeys power-law form, that is $p(k) \sim k^{-\gamma}$, where $k$
is the degree and $p(k)$ is the probability density function for
the degree distribution. $\gamma$ is called the power-law
exponent, and usually between 2 and 3 in real
world\cite{Reviews1,Reviews2,Reviews3}. This power-law
distribution falls off much more gradually than an exponential
one, allowing for a few nodes of very large degree to exist.
Networks with power-law degree distribution are referred to as
scale-free networks, although one can and usually does have scales
present in other network properties.

Synchronization is observed in a variety of natural, social,
physical and biological systems\cite{Phe1,Phe2,Phe3,Phe4}. To
understand how the network structure affects the synchronizability
of the network is of not only theoretical interest, but also
practical value. There are many previous studies about collective
synchronization, with a basic assumption that the dynamic system
of coupled oscillators evolves either on regular
networks\cite{Reg1,Reg2,Reg3}, or on random ones\cite{Reg4,Reg5}.
Very recently, scientist focus on synchronization on complex
networks, and find that the networks of small-world effect and
scale-free property may be easier to synchronize than regular
lattices\cite{syn3,syn4,syn5,syn6,syn8,syn10}.

Since there are countless topological characters for networks, a
natural question is addressed: what is the most important factor
by which the synchroizability of the networks is mainly
determined? some previous works indicated the average distance $L$
is one of the key factors, the smaller $L$ will lead to better
synchronizability\cite{syn4,syn5,syn8}. Other researchers focus on
the role degree heterogeneity plays. They found greater
heterogeneity will result in poorer synchronizability, and
demonstrated that the maximal betweenness\cite{ex3,betweenness1}
$B_{m\texttt{ax}}$ may be a proper quantity to estimate the
network synchronizability. With smaller $B_{m\texttt{ax}}$, the
network synchronizability will be
better\cite{betweenness2,betweenness3}. However, the above results
and conclusions are still debated.

In this paper, we investigate the collective synchronization of
system of coupled oscillators on Barab\'{a}si-Albert scale-free
networks(BA networks for short)\cite{SFN2}. We propose an approach
of structural perturbations, which can markedly enhance the
network synchronizability, and is easy to be realized. It also
provides a theoretical evidence that the maximal betweenness plays
a main role in network synchronization.

This paper is organized as follow: in section 2, the conception of
synchronizability will be briefly introduced. In section 3, we
will describe the approach of structural perturbations. Next, the
simulation results will be given. Finally, in section 5, the
conclusion is drawn, and the relevance of this approach to some
real-life problems is discussed.
\begin{figure}
\scalebox{0.65}[0.65]{\includegraphics{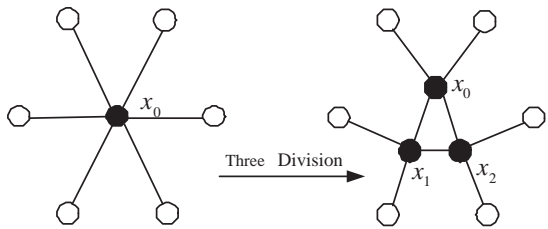}} \caption{The
sketch maps for 3-division  process on $x_0$. The solid circle in
left side is the node $x_0$ with degree 6. After 3-division
process, the $x_0$ is divided into 3 nodes $x_0, x_1$ and $x_2$
that are fully connected. The six edges incident from $x_0$
redistribute over these three nodes.}
\end{figure}

\section{Synchronizability}
We introduce a generic model of coupled oscillators on networks
and the master stability function\cite{master1}, which is often
used to test the stability of the complete synchronized stats.

Each node of a network is located an oscillator, a link connecting
two nodes represents coupling between the two oscillators. The
state of the $i$th oscillator is described by $\textbf{x}^i$, we
get the set of equations of motion governing the dynamics of the
$N$ coupled oscillators:
\begin{equation}
\dot{\textbf{x}}^i=\textbf{F}(\textbf{x}^i)+\sigma\sum_{j=1}^NG_{ij}\textbf{H}(\textbf{x}^j),
\end{equation}
where $\dot{\textbf{x}}^i=\textbf{F}(\textbf{x}^i)$ governs the
dynamics of individual oscillator, $\textbf{H}(\textbf{x}^j)$ is
the output function and $\sigma$ the coupling strength. The
$N\times N$ coupling matrix $\textbf{G}$ is given by
\begin{equation}
    G_{ij}=\left\{
    \begin{array}{cc}
    -k_i   &\mbox{for $i=j$}\\
     1    &\mbox{for $j\in\Lambda_i$}   .\\
     0    &\mbox{otherwise}
    \end{array}
    \right.
\end{equation}
All the eigenvalues of matrix $\textbf{G}$ are nonpositive reals
because $\textbf{G}$ are negative semidefinite, and the biggest
eigenvalue $\gamma_0$ is always zero because the rows of
$\textbf{G}$ have zero sum. Thus, the eigenvalues can be ranked as
$\gamma_0>\gamma_1>\cdots>\gamma_{N-1}$, and the synchronization
manifold is an invariant manifold, that is, the fully synchronized
state: $\textbf{x}^1=\textbf{x}^2=\cdots=\textbf{x}^N=\textbf{s}$
satisfies $\dot{\textbf{s}}=\textbf{F}(\textbf{s})$.

Let $\xi^i$ be the variation on the $i$th node, and the collection
of variation is $\xi=(\xi^1,\xi^2,\cdots=\xi^N)$, we get the
variational equation of (1)
\begin{equation}
\dot{\xi}=[\textbf{1}_N\otimes D\textbf{F}+\sigma\textbf{G}\otimes
D\textbf{H}]\xi.
\end{equation}
Diagonalizing $\textbf{G}$ in the second term of (3), a block
diagonalized variational equation is obtained and each block has
the form
\begin{equation}
\dot{\xi}_k=[D\textbf{F}+\sigma\gamma_k D\textbf{H}]\xi_k,
\end{equation}
where $D$ denotes the Jacobian matrix, $\gamma_k$ is an eigenvalue
of $\textbf{G}$, $k=0,1,2,\cdots,N-1$. $k=0$ corresponds to the
mode that is parallel to the synchronization manifold. Let
$\alpha=\sigma\gamma_k$, rewrite (4) as
\begin{equation}
\dot{\zeta}=[D\textbf{F}+\alpha D\textbf{H}]\zeta.
\end{equation}
Since $D\textbf{F}$ and $D\textbf{H}$ are the same for each block,
the largest Lyapunov exponent $\lambda_{\texttt{max}}$ of Equ. (5)
only depends on $\alpha$. The function
$\lambda_{\texttt{max}}(\alpha)$ is named master stability
function, whose sign indicates the stability of the mode: the
synchronized state is stable if $\lambda_{\texttt{max}}(\alpha)<0$
for all blocks.

For many dynamical systems, the master stability function is
negative in a single, finite interval $(\alpha_1,\alpha_2)$ the
largest Lyapunov exponent is negtive\cite{master2}. Therefore, the
network is synchronizable for some $\sigma$ when the eigenratio
$r=\gamma_{N-1}/\gamma_1$ satisfies
\begin{equation}
r<\alpha_2/\alpha_1
\end{equation}
The right-hand side of this equation depends only on the dynamics
of individual oscillator and the output function, while the
eigenratio $r$ depends only on the coupling matrix $\textbf{G}$.
The problem of synchronization is then divided into two parts:
choosing suitable parameters of dynamics to broaden the interval
$(\alpha_1,\alpha_2)$ and the analysis of eigenratio of the
coupling matrix. The eigenratio $r$ of coupling matrix indicates
the synchronizability of the network, the smaller it is the better
the synchronizability and vice versa. In this paper, for
universality, we will not address a particular dynamical system,
but concentrate on how the network topology affects eigenratio
$r$.

\section{Structural Perturbations}
As mentioned above, the nodes with very large betweenness, namly
hubs, may reduce the network synchronizability. So the present
method of structural perturbations aims at these hubs. For a hub
$x_0$, we add $m-1$ assistant nodes around it, labelled by
$x_1,x_2,\cdots,x_{m-1}$. These $m$ nodes are fully connected.
Then, all the edges incident from $x_0$ will relink to a random
picked node $x_i(i=0,1,\cdots,m-1)$. After this process, the
betweenness of $x_0$ is divided into $m$ almost equal parts
associating with those $m$ nodes. We call this process
$m$-division for short. A sketch map of a 3-division process on
node $x_0$ is shown in Fig. 1.

Due to the huge size of many real-life networks, it is usually
impossible to obtain the nodes' betweenness. Fortunately, previous
studies showed that there exists strongly positive correlation
between degree and betweenness in BA networks and some other real
heterogeneity networks\cite{load1,load2}, that is to say, the node
with larger degree will statistically have higher betweenness.
Therefore, for practical reason, we assume the node with higher
betweenness is surely of larger degree in BA networks. So
hereinafter, all the judgements and operations are based on the
degree of nodes.

In order to enhance the network synchronizability, a few nodes
with highest degree will be divided. Rank each node of a give
network $G$ according to its degree, the node that has highest
degree is arranged in the top of the queue. Then, the network
$G(\rho,m)$ can be obtained by the following $N\rho$ steps. First,
carry out $m$-division on the top node in $G$, leading to the
network $G(\frac{1}{N},m)$. secondly, calculate all nodes' degree
in $G(\frac{1}{N},m)$, and rank each node according to it. Then,
get the network $G(\frac{2}{N},m)$ by dividing the top node in
$G(\frac{1}{N},m)$. Repeat this process $N\rho$ times, with
$N\rho$ nodes have been divided in total, one will reach the
network $G(\rho,m)$. Since the randomicity is involved in dividing
process, $G(\rho,m)$ is not unique. In this letter, we focus on
the difference between $G(\rho,m)$ and $G$.

\section{Simulations}
To explore how the structural perturbations affect network
synchronizability, we compare the eigenratio $r$ before and after
dividing processes. The BA networks of size  $N=2000$ and average
degree $\langle k \rangle=12$ are used for simulations. In Fig. 2,
we report the ratio $R=r'/r$ against the number of nodes that are
divided, where $r$ is the eigenratio of the original network and
$r'$ after the operation. Here we set $m=3$. With the probability
$\rho$ or the number of divided nodes increasing, the ratio $R$ is
observed to decrease, indicating the enhancement of
synchronizability. In Fig.2, it can be seen that to divide a very
few nodes will sharply enhance the network synchronizability. $R$
decreases to 0.7 when only 5 nodes are divided, and will drop to
half after 0.6\% nodes (i.e. 12 nodes) are divided.
\begin{figure}
\scalebox{0.8}[0.8]{\includegraphics{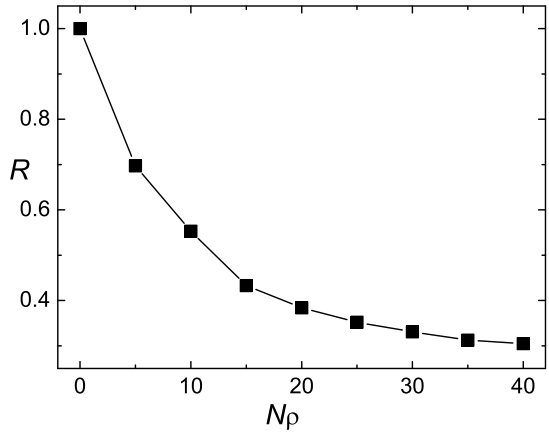}} \caption{Behavior of
the ratio of the eigenratio of network $G(N\rho,m)$ to that of
network $G$ versus the number of divided nodes. As the number
increases, the ratio is shown to reduce, leading to better
synchronization. The average is taken over 50 different network
realizations.}
\end{figure}
\begin{figure}
\scalebox{0.84}[0.8]{\includegraphics{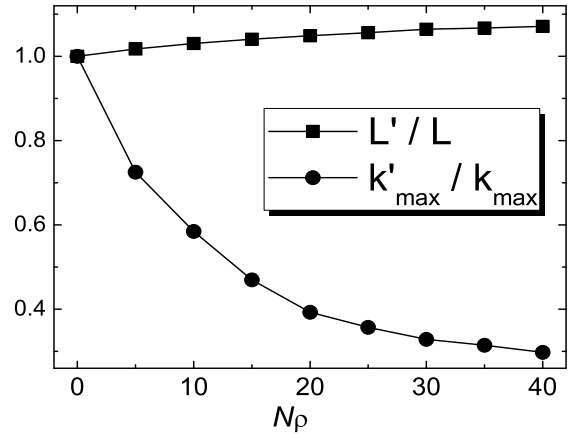}} \caption{The
average distance $L'$ and maximal degree $k'_{\texttt{max}}$ in
$G(\rho,m)$ vs $\rho$. $L$ and $k_{\texttt{max}}$ denote the
average distance and maximal degree in original network $G$. We
plot the relative changes $\frac{L'}{L}$ and
$\frac{k'_{\texttt{max}}}{k_{\texttt{max}}}$ using squares and
circles, respectively. One could see clearly, the dividing
processes reduce the maximal degree while increase the average
distance. All the data are obtained by the average over 20
independent runs.}
\end{figure}
To better understand the underlying mechanism of synchronization
and the reason why these structural perturbations will greatly
enhance network synchronizability, we investigate the behaviors of
two extensively studied quantities, the average distance $L$ and
maximal degree $k_{\texttt{max}}$. In BA networks, the node with
maximal degree is most probably the very node having maximal
betweenness. As illustrated in Fig. 3, $L$ will increase with
$\rho$, while $k_{\texttt{max}}$ will decrease. This results
provide some evidences how the two factors affect the
synchronization of systems. The maximal degree(i.e. the maximal
betweenness) may play the main role in determining network
synchronizability. It is worthwhile to emphasize that from the
simulation results, we can not say anything about how the average
distance affects the network synchronizability. $L$ varies
slightly, and probably has nonsignificant influence compared with
the change of $k_{\texttt{max}}$. This results suggest that to
reduce the maximal betweenness of networks is a practical and
effective approach to enhance the network synchronizability.

\section{Conclusion and Discussion}
Motivated by the practical requirement and theoretical interest,
numbers of scientists begin to study how to enhance the network
synchronizability, especially for scale-free
networks\cite{enhance1,enhance2}. These methods keep the network
topology unchanged, while add some weight into the system, thus
the coupling matrix is changed. These excellent works do not need
any new nodes, new edges or rewirement, but highly enhance the
network synchronizability. However, it may require more
intelligent system, in which each node at least should remember
its particular linking strengths. In this paper, we propose an
approach to enhance the network synchronizability. This approach
does not require any intelligence of nodes, but the network
structure will be slightly changed. In some real-life
communication networks such as Internet, the long length edge may
cost much more than node and short length edge\cite{cost1,cost2},
so if all the nodes $x_1,x_2,\cdots,x_{m-1}$ are in $x_0$'s
vicinal locations, our method is feasible.

Some recent works about network traffic dynamics reveal that the
communication ability of network, called network
throughput\cite{Traffic2,Traffic3}, is mainly determined by the
maximal betweenness, thus to steer clear of those hub nodes may
enhance the network throughput\cite{Traffic3,Traffic4}. This is
just the case of network synchronization. Some method that can
enhance the network throughput will enhance the network
synchronizability too\cite{Traffic2,Traffic4,enhance1,enhance2}.
Therefore, we guess there may exist some common features between
network traffic and network synchronization, although they seem
completely irrelevant. We believe our work will enlighten readers
on this object, and is also relevance to traffic control on
networks.

\begin{acknowledgments}
This work is supported by the National Natural Science Foundation
of China under Grant No. 10472116, 70471033 and 70271070, and the
Specialized Research Fund for the Doctoral Program of Higher
Education (SRFDP No.20020358009).
\end{acknowledgments}


\begin{thebibliography} {Albert2000}
\bibitem{Reviews1} R. Albert and A. -L. Barab\'{a}si, Rev. Mod. Phys. {\bf 74}, 47(2002).
\bibitem{Reviews2} S. N. Dorogovtsev and J. F. F. Mendes, Adv. Phys. {\bf 51}, 1079(2002).
\bibitem{Reviews3} M. E. J. Newman, SIAM Review {\bf 45}, 167(2003).

\bibitem{Epidemic1} R. Pastor-Satorras and A. Vespignani, Phys. Rev, Lett. {\bf 86}, 3200(2001).
\bibitem{Epidemic2} G. Yan, T. Zhou, J. Wang, Z. -Q. Fu and B. -H. Wang, Chin. Phys. Lett. {\bf 22}, 510(2005).
\bibitem{Epidemic3} T. Zhou, G. Yan and B. -H. Wang, Phys. Rev. E {\bf 71}, 046141(2005).
\bibitem{Cascade1} A. E. Motter and Y. -C. Lai, Phys. Rev. E {\bf 66}, 065102(2002).
\bibitem{Cascade2} K. -I. Goh, D. -S. Lee, B. Kahng and D. Kim, Phys. Rev. Lett. {\bf 91}, 148701(2003).
\bibitem{Cascade3} T. Zhou and B. -H. Wang, Chin. Phys. Lett. {\bf 22}, 1072(2005).
\bibitem{Traffic1} B. Tadi\'{c}, Phys. Rev. E {\bf 69}, 036102(2004).
\bibitem{Traffic2} L. Zhao, Y. -C. Lai, K. Park and N. Ye, Phys. Rev. E {\bf 71}, 026125(2005).
\bibitem{Traffic3} G. Yan, T. Zhou, B. Hu, Z. -Q. Fu and B. -H. Wang, arXiv: cond-mat/0505366.
\bibitem{SWN1} D. J. Watts and S. H. Strogatz, Nature {\bf 393}, 440(1998).
\bibitem{SWN3} C. -P. Zhu, S. -J. Xiong, Y. -J. Tian, N. Li and K. -S. Jiang, Phys. Rev. Lett. {\bf 92}, 218702(2004).
\bibitem{SFN2} A. -L. Barab\'{a}si and R. Albert, Science {\bf 286}, 509(1999).
\bibitem{ex1} In a network, the distance between two nodes is
defined as the number of edges along the shortest path connecting
them. The average distance $L$ of the network, then, is defined as
the mean distance between two nodes, averaged over all pairs of
nodes.
\bibitem{ex2} The clustering coefficient $C$ denotes the probability
that randomly picked two neighbors of a random selected node are
neighbors.
\bibitem{Phe1} S. H. Strogatz and I. Stewart, Sci. Am. {\bf 269}, 102(1993).
\bibitem{Phe2} C. M. Gray, J. Comput. Neurosci. {\bf 1}, 11(1994).
\bibitem{Phe3} L. Glass, Nature, {\bf 410}, 277(2001).
\bibitem{Phe4} Z. N\'{e}da , E. Ravasz, T. Vicsek, A. L. Barab\'{a}si, Phys. Rev. E {\bf 61}, 6987(2000).
\bibitem{Reg1} J. F. Heagy, T. L. Carroll and L. M. Pecora, Phys. Rev. E {\bf 50}, 1874(1994).
\bibitem{Reg2} C. W. Wu and L. O. Chua, IEEE Trans. Circuits Syst. I {\bf 42}, 430(1995).
\bibitem{Reg3} J. Jost and M. P. Joy, Phys. Rev. E {\bf 65}, 016201(2001).
\bibitem{Reg4} P. M. Gade, Phys. Rev. E {\bf 54}, 64(1996).
\bibitem{Reg5} S. C. Manrubia and A. S. Mikhailov, Phys. Rev. E {\bf 60}, 1579(1999).
\bibitem{syn3} L. F. Lago-Fern¨¢ndez, R. Huerta, F. Corbacho, and J. A. Siguenza, Phys. Rev. Lett. {\bf 84}, 2758 (2000).
\bibitem{syn4} P. M. Gade and C. -K. Hu, Phys. Rev. E {\bf 62}, 6409 (2000).
\bibitem{syn5} X. F. Wang and G. Chen, Int. J. Bifurcation Chaos Appl. Sci. Eng. {\bf 12}, 187(2002).
\bibitem{syn6} P. -Q. Jiang, B. -H. Wang, S. -L. Bu, Q. -H. Xia, and X. -S. Luo. Int. J. Mod. Phys. B {\bf 18}, 2674(2004).
\bibitem{syn8} M. Barahona and L. M. Pecora, Phys. Rev. Lett. {\bf 89}, 054101(2002).
\bibitem{syn10} P. G. Lind, J. A. C. Gallas, and H. J. Herrmann, Phys. Rev. E {\bf 70}, 056207 (2004).
\bibitem{ex3} The betweenness of a given node denotes the
probability that a shortest path with random source and
destination will pass through this node. One would see ref.
\cite{betweenness1} for details.
\bibitem{betweenness1} M. E. J. Newman, Phys. Rev. E {\bf 64}, 016131(2001).
\bibitem{betweenness2} T. Nishikawa, A. E. Motter, Y. -C. Lai, and F. C. Hoppensteadt, Phys. Rev. Lett. {\bf 91}, 014101(2003).
\bibitem{betweenness3} H. Hong, B. J. Kim, M. Y. Choi, and H. Park, Phys. Rev. E {\bf 69}, 067105(2004).
\bibitem{master1} L. M. Pecora and T. L. Carroll, Phys. Rev. Lett. {\bf 80}, 2109(1998).
\bibitem{master2} G. Hu, J. Yang, and W. Liu, Phys. Rev. E {\bf58}, 4440(1998).
\bibitem{load1} K. -I. Goh, B. Kahng, and D. Kim, Phys. Rev. Lett. {\bf 87}, 278701(2001).
\bibitem{load2} M. Barth\'{e}lemy, Eruo. Phys. J. B {\bf 38}, 163(2004).
\bibitem{enhance1} A. E. Motter, C. Zhou, and J. Kurths, Phys. Rev. E {\bf71}, 016116(2005).
\bibitem{enhance2} M.Chavez, D. -U. Hwang, A. Amann, H. G. E. Hentschel, and Boccaletti, Phys. Rev. Lett. {\bf 94}, 218701(2005).
\bibitem{cost1} B. Waxman, IEEE J. Selec. Areas Commun. {\bf 6}, 1617(1998).
\bibitem{cost2} S. -H. Yook, H. Jeong, and A. -L. Barab\'{a}si, PNAS {\bf 99}, 13382(2002).
\bibitem{Traffic4} C. -Y. Yin, B. -H. Wang, W. -X. Wang, T. Zhou, and
H. -J. Yang, arXiv: physics/0506204.




\end{thebibliography}
\end{document}